\documentclass[11pt, oneside]{article}   	
\usepackage{geometry}                		
\usepackage{graphicx}				
\usepackage{amssymb}
\usepackage[usenames,dvipsnames]{color,xcolor}

\title{Axiomatic conformal  theory in dimensions $>2 $ and AdS/CT  correspondence}
\author{Albert Schwarz}

\date{}							

\begin{document}

{\maketitle
}
{\it To Sasha Polyakov with admiration and love}
\vskip .1in
\abstract { We formulate axioms of conformal theory (CT) in dimensions $>2$ modifying Segal's axioms for two-dimensional CFT. (In the definition of higher-dimensional CFT  one includes also a condition of existence of energy-momentum tensor.) We use these axioms to derive 
the AdS/CT correspondence for local theories on AdS . We introduce a notion of weakly local quantum field theory and  construct a bijective correspondence between  conformal  theories on the sphere $S^d$ and weakly local  quantum field theories on $H^{d+1}$ that are invariant with respect to isometries. (Here $H^{d+1}$ denotes hyperbolic space= Euclidean AdS space.) We give an expression of AdS correlation functions in terms of CT correlation functions.
The conformal  theory has conserved energy-momentum tensor iff the AdS theory has graviton in its spectrum. }


\section {Introduction}

The AdS/CFT correspondence \cite {1}, \cite {2}, \cite {3} played very important role in the development of quantum field theory and string theory. This correspondence relates string theory on AdS with conformal field theory on the boundary. It was understood very soon \cite {2}, \cite {3} that a similar correspondence can be constructed for local quantum theories on AdS and conformal theories (CT) on the boundary (AdS/CT correspondence).
The main goal of this paper is to give a very simple  rigorous proof of  the  AdS/CT correspondence for local theories. We show that,  for every local quantum field theory on  $(d+1)$-dimensional AdS that is invariant with respect to isometries,  one can construct $d$-dimensional conformal field theory with the same space of states. Moreover, our construction can be applied also to weakly local theories  (see Section 3 for the definition of weak locality.) The CT has  a conserved energy-momentum tensor iff the theory on AdS has  the  graviton in its spectrum. (Notice, however, that our constructions can be applied to quantum gravity only in the framework of perturbation theory.)

Let us emphasize that  our statement does not cover  the original example of $N=4$ SYM theory that comes from string theory ( not from local quantum field theory). 

 We did not analyze in detail the relation of our considerations to the existing heuristic constructions (see \cite {KAP}, \cite {RAM}, \cite {HAR} for review). It seems these constructions do not always lead to genuine conformal theories (Polyakov, private communication); in those cases they definitely differ from our construction.  It is clear, however, that  our  formulas either  agree with standard constructions, or constitute a more precise version of these constructions.\footnote { A rigorous proof of AdS/CT correspondence was claimed in \cite {REH}. However, it seems that Rehren's construction not   necessarily leads to  conformal theories with OPE.}

We work in the Euclidean setting. Hence our AdS is Euclidean AdS that is hyperbolic space 
(Lobachevsky space) from the viewpoint of mathematician and our conformal theories are defined on $S^d$ or $\mathbb{R}^d.$

Our proof is based on the axiomatics of  conformal theory in dimensions  $>2$. Our axioms modify Segal's axioms for two-dimensional CFT \cite {S1},\cite {S2}. ( Segal's papers contain also discussion of axioms of quantum field theory in the general case.) Segal starts with Riemann surfaces (two-dimensional conformal manifolds) having  holes with parameterized boundaries. To every boundary he assigns vector space $H$. The holes are divided in two classes ("incoming" and "outgoing''). {\footnote { Segal talks about cobordisms instead of incoming and outgoing holes, but this is only terminological difference.}} If we have $m$ incoming holes and $n$ outgoing holes  CFT specifies a map $H^{\otimes m}\to H^{\otimes n}.$ Segal's axioms describe what happens if we sew two surfaces. Our axioms for higher-dimensional theories are based on the same ideas. We consider the standard $S^d$ of radius 1 with holes, but we allow only round holes. We do not  consider two  types of holes, but this is irrelevant. We could modify our axioms to consider both types of holes. Instead of talking about sphere with holes we are talking about collections of non-overlapping parameterized round balls. The conformal group acts on these collections; factorizing the space of collections with respect to this action,  we obtain the  space $\mathcal {M}_n$, an analog of moduli space of Riemann surfaces with holes in our setting.  Notice that $\mathcal {M}_n$ is finite-dimensional;  this is related to the fact that the conformal group is finite-dimensional in dimensions $>2. $ To specify conformal theory (CT)  we assign to every element of $\mathcal {M}_n$  an $n$-linear functional on the space of states $H$
(an element of  a tensor power of  $H^*$). We formulate axioms of CT and analyze their relation   to other approaches.  Following the suggestion of  \cite {KAP} we reserve the name CFT for CT with conserved energy-momentum tensor.

Axiomatic conformal field theory became very fashionable recently under the funny name "conformal bootstrap". The renewed interest to 
conformal bootstrap suggested by  A. Polyakov many years ago was generated by papers  where it was shown that the axioms of unitary CFT are strong enough to prove very good estimates for anomalous dimensions in 3D Ising model \cite {R}, \cite {RR}.  

To derive the AdS/CT correspondence,  we notice that one can construct the space 
$\mathcal {M}_n$ starting with hyperbolic space $H^{d+1}$ (we should
consider half-spaces instead of balls). Now having a local quantum field theory on hyperbolic space we can define functionals entering the definition of CT. (If the theory is determined by  a local action $S$, we integrate $e^{-S}$ over the complement to half-spaces.)

The paper  does not depend on any papers about CFT  or about AdS. In Section 2 we formulate our axioms of CT and in Section 4 we relate them  to other approaches. In Section 3 we derive  the  AdS/CT correspondence. In Section 5 we discuss the  AdS/CT dictionary. In particular, we express AdS partition functions and AdS correlation functions in terms of CT correlation functions.  It is not clear whether our dictionary is completely equivalent to existing ones; however, we show that it is very close both to GKPW  
dictionary suggested in \cite {2}, \cite {3} and to  BDHM dictionary suggested in \cite {BD} (see \cite {HAR}, \cite {KAP} for review).
 
\section {Axiomatic conformal  theory}

The  group of conformal transformations of the sphere $S^d$  is denoted by ${\rm Conf}_d$; it is generated by inversions. Its connected component is isomorphic to $ SO(1, d+1)$. We define a round ball in $S^d$ as a conformal map of the standard round ball into $S^d.$  Notice that this means that we have fixed a conformal parameterization of the boundary of a round ball in $S^d$ (a conformal map of $S^{d-1}$ onto the boundary ).
Let us consider the space  of $n$ non-overlapping round balls on the sphere $S^d$. The conformal transformations act on this space; we denote  by $\mathcal{M}_n $ the space of conformal classes of ordered collections of $n$  non-overlapping round balls (the space of orbits of  ${\rm Conf}_d$  in the space of collections of balls).  The sphere $S^d$ is conformally equivalent to the Euclidean space $\mathbb{R}^d$; round balls in $S^d$ correspond to round balls, complements to round balls  and half-spaces in $\mathbb{R}^d $  with conformal parameterization of boundaries. The  space $\mathcal{M}_1$ consists of one point, in general the space $\mathcal{M}_n$ is  a smooth manifold of  dimension $(n-1) \dim {\rm Conf}_d=\frac{(n-1)(d+2)(d+1)}{2}. $ The group of permutations $S_n$ acts on $\mathcal{M}_n$ in an obvious way. One can construct a natural map $\phi_{nm}: \mathcal{M}_n\times \mathcal{M}_m\to \mathcal{M}_{n+m-2}$. To construct this map we will work  in $\mathbb{R}^d.$ Then performing a conformal transformation  we can consider the last ball in $\mathcal{M}_n$ as the half-space $x_d\geq 0$ and the first ball in$\mathcal{M}_m$ as the half-space $x_d\leq 0.$  The remaining $m+n-2$ balls specify  a point in $\mathcal{M}_{m+n-2}.$ ( We can represent a ball as a half-space in many ways. However, we have fixed a conformal parameterization of the ball ; this allows us to specify a unique transformation of the ball onto half-space.) \footnote { Our construction is reminiscent  of  the definition of little disks operad.} 
Notice that the map $\phi_2$ specifies an associative multiplication on $\mathcal{M}_2$; in other words $\mathcal{M}_2$ can be considered as semigroup. More generally, the operations $\phi_{nm}$ specify associative multiplication in the union $\mathcal{M}$ of spaces $\mathcal{M}_n.$  The map $\phi_{n,2}$ determines an action of the semigroup $ \mathcal{M}_2$ on $\mathcal{M}_n.$

Of course, the construction of the map $\phi _{mn}$ can be given directly in $S^d.$ In  particular, the action of the semigroup $ \mathcal{M}_2$ on $\mathcal{M}_n$ replaces the last ball in the collection specifying an element of action of the semigroup $ \mathcal{M}_2$ on $\mathcal{M}_n$ by a smaller ball in the interior of the last ball.

To  give an axiomatic description of CT  we fix a topological vector space $\mathcal{H}$ (the space of states) and an element $a\in  \mathcal{H}\otimes \mathcal{H}.$  In a basis $e_i$ of $\mathcal {H}$ we can write  $a= a^{ik}e_ie_k$. The element $a$  determines  an associative multiplication in the direct sum $H$ of vector spaces $(\mathcal{H}^*)^{\otimes n}$ dual to tensor  powers $\mathcal{H}^{\otimes n}$. \footnote {Sometimes it is convenient to consider instead of $\mathcal{H}^*$ a dense subspace of it. We will disregard these subtleties.} In  the basis $e_i$ the elements of $H$ can be represented as  covariant tensors of various ranks. We can represent the product of a  tensor $r_{i_1,...,i_n}$  (= a linear functional  on $\mathcal{H}^{\otimes n}$ ) and a  tensor  $s_{k_1,...,k_m}$ (= a linear functional on $\mathcal{H}^{\otimes m}$) as a tensor of rank $n+m-2$ (= a linear  functional on $\mathcal{H}^{\otimes (n+m-2)}$)   as a contraction of the last index of $r$ with the first index of $s$ by means of the tensor $a^{ik}$.  Notice that the tensor $a$ specifies an inner product in $\mathcal{ H}^*$; the multiplication can be defined in terms of this product.

{ \it We assume that for every point of} $\mathcal{M}_n$ {\it we have a  map} $\psi_n:\mathcal{H}^{\otimes n}\to \mathbb{C}$  ( a  multilinear functional $\psi_n(h_1,...,h_n)$ where $h_k\in \mathcal{H})$. This functional should depend continuously  on the point of $\mathcal{M}_n$. If necessary to emphasize the dependence on the point of $\mathcal{M}_n$ we will use the notation $\psi_n(B_1,...,B_n, h_1,..., h_n)$ where $B_1, .., B_n$ are balls specifying this point. Together the functionals $\psi_n$  determine  a continuous map $\Psi : \mathcal{M}\to H.$  We assume that this map commutes with the actions of the   group of permutations $S_n$ , i.e. {\it the functional $\psi_n(B_1,...,B_n, h_1,..., h_n)$ is $S_n$-invariant.}{\it The main axiom of} CT {\it  is the requirement that the map $\Psi$ is a homomorphism} (the product in $\mathcal{M}$ goes to the product in $H$).\footnote { In two-dimensional theories the infinite-dimensional conformal Lie algebra has central extension, therefore we should allow projective representations. Conformal Lie algebra in dimension $>2$ does not have central extensions, but still it is possible that the homomorphism $\Psi$ is multivalued.}
 
 One can reformulate the main axiom in the following way. Let us consider non-overlapping balls $B_1,..., B_{r+s}$  specifying an element of $\mathcal{M}_{r+s}$ and corresponding functional
$\psi_{r+s} (h_1,...h_{r+s}).$ Let us choose a sphere $S^{d-1}$ in such a way that the first $r$ balls are inside the sphere and the last $s$ balls are outside it.
This sphere bounds two balls $B_{in}$ and $B_{out}.$ The balls $B_1, ..., B_r, B_{out}$ specify an element of $\mathcal {M}_{r+1}$. For fixed $h_1,...h_r$ the corresponding functional $\psi _{r+1}$  determines an element $\Psi_1=
\Psi_1(h_1,...,h_r)\in\mathcal{ H}^*$. The balls $B_{in},B_{r+1},..., B_{r+s}$ specify an element of $\mathcal {M}_{s+1}$. For fixed $h_{r+1},...h_{r+s}$ the corresponding functional $\psi _{s+1}$  determines an element $\Psi_2=\Psi_2 (h_{r+1},...h_{r+s})\in \mathcal{ H}^*.$ An equivalent formulation of the main axiom is the expression of $\psi_{r+s}$ as the inner product of $\Psi_1$ and $\Psi_2$ :\begin{equation}
\label{2m}
\psi_{r+s} (h_1,...h_{r+s})=<\Psi_1(h_1,...,h_r),\Psi_2 (h_{r+1},...h_{r+s}>.
\end{equation}
(Recall that the tensor $a$ specifies an inner product in $\mathcal{ H}^*.$)

 Let us explain the physical origin of these constructions. Let us consider a conformally invariant local action functional $\mathcal{S}$ on $\mathbb{R}^d$  or, equivalently, on  $S^d.$ Let us    calculate the corresponding partition function  on the domain $V_n$ obtained from $S^d$ by deleting $n$ balls as a functional integral of $e^{- \mathcal{S}}$  over the space of fields on $V_n.$ 
 This partition function depends on 
 s; it should be identified with $\psi_n(h_1,..., h_n)$. (Hence $\mathcal{H}$ should be identified with the space of boundary states.)  The main axiom of CT comes from  the remark
 that $V_{n+m-2}$ can be represented as a union of $V_m$ and $V_n$ having a common part of boundary that can be identified with $S^{d-1}.$ 
(To  calculate $\psi_{n+m-2}$ we do the integral over fields defined on $V_{n+m-2}.$ We can do this in two steps. First, we calculate the integrals over  the fields defined on $V_n$ and $V_m,$ we get $\psi_n$ and $\psi_m$.
 Second, we paste together these two answers inserting a $\delta$-function that guarantees that the fields on $V_n$ and $V_m$  coincide on the 
  common boundary    and integrating over the fields on this boundary. This integration gives us a scalar product on the space $\mathcal {H}^*$.)\footnote {Notice, that our considerations did not use conformal invariance in any way, they were based only on locality of action. Moreover, even locality is not quite necessary; see below.}

Let us consider the homomorphism  $\psi _2: \mathcal{M}_2\to \mathcal{H}^*\otimes\mathcal{H}^*$ in more detail.  The multiplication in the space $\mathcal{H}^*\otimes\mathcal{H}^*$ can be represented in coordinates as an operation transforming a pair of tensors $x_{ik}, y_{ik}$ into the tensor $z_{ik}=x_{il}a^{ls}y_{sk}. $ Raising the second index of tensor $x_{ik}$ by means of tensor $a^{kl}$ we obtain a tensor $\tilde x^s_i=x_{il}a^{ls},$ that can be considered as an element of  the ring $End \mathcal{H}$ of linear operators in $\mathcal{H}.$ It is easy to check $\tilde z_i^k=\tilde x_i^s\tilde y_s^k.$ This means that $\psi_2$ specifies a homomorphism  of $ \mathcal{M}_2\to End \mathcal{H}$ . In other words, the semigroup $\mathcal{M}_2$ acts on $\mathcal{H}.$ It is easy to verify that the Lie algebra of the semigroup  $\mathcal{M}_2$ coincides with the Lie algebra
$so(1, d+1)$ of the group $SO(1, d+1)$. (To prove this fact we notice that  in $\mathbb{R}^d$ every element of $\mathcal{M}_2$ can be represented as  the exterior of the unit ball and a parameterized round ball inside the unit ball. This representation is unique. This remark allows us to identify $\mathcal{M}_2$ with the subsemigroup of ${\rm Conf}_d$ that consists of elements mapping  the unit ball into its interior.) We conclude that  this Lie algebra acts on $\mathcal H.$ An important one-dimensional subsemigroup $\mathcal{L}$ of $\mathcal{M}_2$ corresponds to dilations. An element of $\mathcal{L}$
consists of two balls having centers in  the south pole and north pole  of $S^d$ respectively (the parameterizations are  fixed in such a way that the corresponding points lie at the same great circle).  In the $\mathbb{R}^d$ picture we should fix some point and consider the interior of a sphere with a center at this point and the exterior of a larger sphere with the same center.
The corresponding element of $\mathcal{L}$ will be denoted by $T_{\alpha}$ where $\alpha=\log \frac{R}{r}$ where $r$ stands for smaller radius, $R$ for  larger radius. It is easy to check that $T_{\alpha}T_{\beta}=T_{\alpha+\beta}.$
The infinitesimal generator of the subgroup $\mathcal{L}$ will be denoted by $S$; we fix this generator in such a way that $T_{\alpha}=e^{-\alpha S}.$  In the Lie algebra of the conformal group ${\rm Conf}_d,$ the element $S$ corresponds to dilation.


\section
{AdS/CT}

To derive the AdS/CT correspondence, we interpret the spaces $\mathcal{M}_n$ in terms of Euclidean AdS space. From the viewpoint of mathematics,  this is the hyperbolic space (Lobachevsky space) $H^{d+1}.$
It can be considered as a connected component of the hyperboloid $x_0^2-x_1^2-...-x^2_{d+1}=R^2$ in $(d+2)$-dimensional space. Equivalently, we can consider the space $\mathbb{R}^{1, d+1}$ with indefinite inner product ( one positive sign and $d+1$ negative signs); then the hyperbolic space is singled out by the equation $<x,x>=R^2$  and inequality $x_0>0.$ ( We will fix $R=1$; in other words we consider hyperbolic space with  curvature $K=-1.$) It follows from this representation that the isometry group of hyperbolic space   is isomorphic to ${\rm Conf}_d$ and its connected component is isomorphic to $SO(1, d+1).$  Applying stereographic projection with the center at the point   $(-1, 0,...0)$, we obtain the Poincar\'e   ball interpretation of hyperbolic space. (We are projecting into the hyperplane $x_0=0$; the hyperbolic space $H^{d+1}$ is identified with the open unit ball $x_1^2+...+x_{d+1}^2<1.$ ) The points of the unit sphere $S^d$ are called boundary points, or ideal points, or points at infinity of the hyperbolic space $H^{d+1}$.  The isometries of $H^{d+1}$ induce conformal transformations on $S^d.$  

Notice that the ideal points of a hyperplane in  $H^{d+1}$ constitute a sphere $S^{d-1}$ conformally embedded into  the ideal sphere $S^d.$  The group  ${\rm Conf}_d$ acts transitively on the space of hyperplanes, hence it  is sufficient to check this statement  for one hyperplane. It is obviously true for the hyperplane $x_1=0$ in the Poincar\'e ball.  Conversely, taking into account that ${\rm Conf}_d$ acts transitively on the space of conformal spheres $S^{d-1} $ in $S^d$, we see that every such sphere consists of ideal points of some hyperplane.  A hyperplane divides  $H^{d+1}$ in two half-spaces; this allows us to analyze ideal points of half-spaces.
 
Let us consider parameterized half-spaces of  $H^{d+1}$ (in other words we consider isometric maps of the standard half-space into hyperbolic space $H^{d+1}$ ).  It follows from the above considerations that parameterized
half-spaces are in one-to-one correspondence with conformally parameterized round balls in $S^d.$ This allows us to describe spaces $\mathcal{M}_n$ in terms of hyperbolic space . Namely, we should consider the space of ordered collections  of $n$  non-overlapping half-spaces $(\Gamma_1,...\Gamma_n)$. The 
group ${\rm Conf}_d$ acts on this space; by definition $\mathcal{M}_n$ is the space of orbits of this action. The definition of associative multiplication in the union $\mathcal{M}$ of the spaces $\mathcal{M}_n$  can be given in the following way. Represent an element of $\mathcal{M}_m$ as a collection of $n$ parameterized half-spaces where the last half-space in the Poincar\'e ball interpretation is $x_1\geq 0$. Represent an element of $\mathcal{M}_n$ as a collection of $m$ parameterized half-spaces where the first half-space in the Poincar\'e ball interpretation is $x_1\leq 0$. Then the first 
$n-1$ half-spaces in the collection of $n$ half-spaces together with last $m-1$ half-spaces in the collection of $m$ half-spaces specify a product of these two elements as an element of
$\mathcal{M}_{n+m-2}.$ 

Now it is easy to prove that a local quantum field theory on  hyperbolic space that is invariant with respect to the isometry group  generates 
$d$-dimensional CT.

 If  such a theory is specified by a local action functional $\mathcal S$, we can construct a partition function  $\psi _n$  that corresponds  to the collection of $n$ half-spaces  $(\Gamma_1,...\Gamma_n)$ by   integrating $e^{-\mathcal S}$ over the fields defined on the complement to the union of half-spaces.  {\it {We assume that this integral makes sense.}}  The partition function depends on the choice of boundary conditions that should be specified on the boundary of every half-space (on   hyperplane) and at infinity ; {\it {we assume that the boundary conditions at   infinity are  ${\rm Conf}_d$-invariant}}. We 
obtain a symmetric functional $\psi_n ( \Gamma_1,...\Gamma_n,h_1, ..., h_n)$ where $h_i$ belongs to the space of boundary states $\mathcal{H}.$ The functionals $\psi_n (h_1, ..., h_n)$  depend on the point of $\mathcal{M}_n$ (because we have assumed that the action is ${\rm Conf}_d$-invariant)  and depend continuously on this point. Together they specify a map $\Psi$ of the space $\mathcal{M}$ into the direct sum $H$ of tensor powers of $\mathcal{H}^*.$  To prove that the ${\rm Conf}_d$-invariant quantum field theory on hyperbolic space $H^{d+1}$ induces CT on $S^d$,  we should  check that this map is a homomorphism. We can do this using  standard manipulations with functional integrals that we repeated already in the case of conformal action functionals. 

Notice that it is not necessary to start with action functionals.  One can  use  an axiomatic definition of local Euclidean QFT on a manifold $X$ that takes as a starting point partition functions $Z_U$ on some domains in $X$ depending on  some data on boundaries of these domains. It is not clear how to formulate full system of axioms for these partition functions (and it seems that some additional data are needed). However, some  requirements are clear. In particular, in the case when two domains $U_1$ and $U_2$ have a common component of  boundary  we should have an expression of  the partition function for $U=U_1\bigcup U_2$ in terms of partition functions for $U_1$ and $U_2.$  For example, let us suppose  that the boundary of $U_1$ has two components $\Sigma_1, \Sigma$ and the boundary of $U_2$ has two components $\Sigma$ and $\Sigma_2$  (here $\Sigma$ is the common component). Then the partition  function $Z_{U_1}$   is a linear functional on the spaces of boundary states, i.e. an element of  
$\mathcal{H}_1^*\otimes \mathcal{H}^*$,  and the partition  function $Z_{U_2}$  is an element of 
$\mathcal{H}\otimes \mathcal{H}_2^*. $   (Notice that the $\Sigma$ enters the boundaries of $U_1$ and $U_2$ with opposite orientations , therefore corresponding spaces of boundary states are dual . Using the pairing between dual spaces we obtain $Z_U$ as an element of $\mathcal{H}_1^*\otimes \mathcal{H}_2^*.$ (Here $\mathcal{H}_i$ stands for boundary conditions on $\Sigma_i.$)  This statement has an obvious generalization to the case of several components of boundary. The generalization  (gluing axiom) can be used to verify that $\Psi$ is a homomorphism.

We have proven that   the ${\rm Conf}_d$-invariant quantum field theory on hyperbolic space $H^{d+1}$ (on Euclidean AdS) induces CT on $S^d.$
 Notice that CT in our definition not necessarily has conserved energy-momentum tensor (is not necessarily  a CFT). We will argue that such a tensor does exist iff the corresponding quantum field theory on hyperbolic space has the  graviton in its spectrum. 
 
 Let us assume now that we have a CT on $S^d$. Can it be obtained from ${\rm Conf}_d$-invariant quantum field theory on hyperbolic space $H^{d+1}$? It is easy to see that for some (non-standard!) definition of quantum field theory the answer is positive. We will say that  a quantum field theory on $H^{d+1}$ is specified by a  symmetric functional $\psi_n ( \Gamma_1,...\Gamma_n,h_1, ..., h_n)$ where $h_i$ belongs to the space of boundary states $\mathcal{H}$ and $\Gamma_i$ are non-overlapping half-spaces; we assume that this functional ( the partition function on the complement to half-spaces $\Gamma_i$) is ${\rm Conf}_d$-invariant . We fix a scalar product on the space $\mathcal{H}^*$. Using this scalar product we can formulate the gluing axiom; if  this axiom is satisfied we say that our quantum field theory is weakly local. \footnote {
The functionals $\psi_n$ specify a map of the union $\mathcal{M}$ of the spaces $\mathcal{M}_n$ into direct sum of vector spaces $\mathcal {H}^{\otimes n}$; the gluing  axiom is equivalent to the statement that this map is a homomorphism with respect to operations described above.} It is obvious that conformal field theories on $S^d$ are in one-to-one correspondence with weakly local ${\rm Conf}_d$-invariant quantum field theory on hyperbolic space $H^{d+1}$ (on Euclidean AdS). 

One can try to apply the above considerations to the string theory on AdS (or on a product of AdS and a  compact manifold). One can consider the partition  function of string on the  domain in AdS bounded by hyperplanes . (Hyperplanes should be considered as $D$-branes or as stacks of $D$-branes. For example, in the case of $AdS_5\times S^5$ one  could consider $D5$-branes of the form $AdS_4\times S^2.$) \footnote {We do not consider $D$-branes as dynamical objects. However, one can formulate an analog of background independence for $D$-branes: a variation of $D$-brane can be represented as a variation of a state on the original $D$-brane.}  It is natural to conjecture  that  the string theory is weakly local; this conjecture is supported by some heuristic considerations. However, this conjecture does not lead to AdS/CFT correspondence; it leads to a particular case of so called  AdS/dCFT correspondence \cite {KR}.

\section
{CT basics}:

We have used an axiomatic approach to CT. Let us discuss the relation of our approach to  standard formalism.  As in the standard approach,  the Lie algebra $so(1,d+1)$ acts on the space of states $\mathcal{H}$. Eigenvectors of the dilation operator $S$ are called scaling states, corresponding eigenvalues are called anomalous dimensions and denoted by $\Delta.$ We assume that scaling states form a basis in $ \mathcal{H}$ (i.e. every element of $\mathcal{H}$ can be presented as a convergent series $\sum c_ne_n$ where $e_n$ are linearly independent scaling states).  Scaling  states  that are highest weight vectors are called primary states.  (Recall that the Lie algebra $so(1,d+1)$  acting on $\mathbb{R}^d$ is generated by translations $P_{\mu}$, othogonal transformations $M_{\mu\nu}$, dilation $S$ and conformal boosts $K_{\mu}$. In these notations, a primary state $\omega$  is  characterized by the condition $K_{\mu}{\omega}=0.$) Every primary state generates a  subrepresentation. Other scaling states  belonging to this subrepresentation are called descendants. One can construct descendants using the remark that for  scaling state  
$\rho$ with anomalous dimension $\Delta$ the state $P_{\mu}\rho$ is a scaling state with anomalous dimension $\Delta +1$. (This follows from the commutation relation $[S,P_{\mu}]=P_{\mu}.$)

To describe correlation functions on $\mathbb{R}^d$ in our approach, 
 we notice  first of all that in the construction  of the action of the semigroup $\mathcal{M}_2$ on $\mathcal{M}_n$ we have singled out the last ball. We can get $n$ actions of $\mathcal{M}_2$ on $\mathcal{M}_n$ adjoining an element of $\mathcal{M}_2$  to other balls. (To get these $n$ actions,  we  can also combine the action we started with  and the action of permutations.) In particular, the direct product of $n$ copies of the semigroup $\mathcal{L}\subset \mathcal{M}_2$ acts on  $\mathcal{M}_n$. This action changes the radii of the balls, but does not change their centers. All these semigroups act also on $\mathcal{H}$; we use the same notation for generators in both cases. By definition,  the functional $\psi_n(B_1, ...,B_n, h_1,...,h_n)$ is compatible with the action of semigroups, in particular
$$\psi_n(e^{-\alpha_1S }B_1,...,e^{-\alpha_nS}B_n,e^{-\alpha_1S}h_1,...,e^{-\alpha_nS}h_n)=\psi_n(B_1, ...,B_n, h_1,...,h_n).$$

Working in $\mathbb{R}^d$  we will introduce notation $B(x,r)$  for the ball of radius $r$ with  center at the point $x$ parameterized in the standard way. Then it follows from the above formula that
\begin{equation}
\label{cf}
\psi_n(B(x_1, 1),...,B(x_n,1),h_1,...,h_n)=\psi_n(B(x_1,r_1),...,B(x_n,r_n), r_1^Sh_1,...,r_n^Sh_n).
\end{equation}
If $h_1,...,h_n$ are scaling states with anomalous dimensions $\Delta_1,..., \Delta_n$ we can rewrite this equation in the form
\begin{equation}
\label{cf2}
\psi_n(B(x_1, 1),...,B(x_n,1),h_1,...,h_n)=\psi_n(B(x_1,r_1),...,B(x_n,r_n), r_1^{\Delta_1}h_1,...,r_n^{\Delta_n}h_n).
\end{equation}
We will use the notation $<\hat h_1(x_1) ...\hat h_n(x_n)>$ for the LHS of (\ref{cf}). Notice 
that the LHS sometimes is not well defined because the unit balls overlap; to define $<\hat h_1(x_1) ...\hat h_n(x_n)>$ in this case we should use  the RHS for small radii $r_i$. It is always well defined in the case when the points
$x_1, ..., x_n$ are distinct. 

In the standard terminology,  the functions $<\hat h_1(x_1) ...\hat h_n(x_n)>$ are correlation functions for local fields $\hat h_i(x)$ corresponding to states $h_i$ in  state -operator correspondence. However, we do not need the notion of local field. Notice that knowing the functions $<\hat h_1(x_1) ...\hat h_n(x_n)>$ and the dilation operator $S,$ we can restore the functions  $\psi_n$ using (\ref {cf}).  The answer is especially simple in the case when $h_i$ are scaling states with anomalous dimensions $\Delta_i$ , then we can use (\ref{cf2}).  We obtain
\begin{equation}
\label{cf3}
\psi_n(B(x_1,r_1),...,B(x_n,r_n), h_1,...,h_n)=r_1^{-\Delta_1}...r_n^{-\Delta_n}<\hat h_1(x_1) ...\hat h_n(x_n)>
\end{equation}
This allows us to derive the axioms we are using  starting with any approach to CT (at least formally). For example, we can start with  the approach of \cite{LM}. From the other side,  one can derive the properties of correlation functions used in other approaches from our axioms. In particular, one can derive the transformation rules for correlation functions from (\ref {cf}) taking infinitesimally small radii in the RHS.

Let us discuss, for example, the derivation of OPE (operator product expansion).
We assume that $h_1,...,h_n$ are scaling states with anomalous dimensions $\Delta_1,..., \Delta_n$ and that the scaling states $e_{\alpha} $ with anomalous dimensions $\Delta _{\alpha}$ form a basis of the space $\mathcal{H}.$ Let us suppose that $||x_2-x_1||<R$ where $R=\min _{i>2}||x_i-x_1||$ .Then there exists a convergent expression
\begin{equation}
\label{ope}
<\hat h_1(x_1) ...\hat h_n(x_n)>=\sum _{\alpha} C_{\alpha}(x_2-x_1)<\hat e_{\alpha} (x_1)\hat h_3(x_3) ...\hat h_n(x_n)>
\end{equation} 
where $C_{\alpha}(x)$ are homogeneous functions of degree $\Delta_1+\Delta_2-\Delta_{\alpha}$ (they depend on states $h_1,h_2, e_{\alpha}$, but do not depend on $h_3,...,h_n.$)
To prove this statement, we apply (\ref {2m}) to  the case when $r=2$,$s=n-2$,  $S^{d-1}$ is a sphere of radius $R-\epsilon$ with the center $x_1$, $B_i$ stands for a small ball with the center at $x_i.$ We decompose the element $\Psi_1$ in a series with respect to the basis $e_{\alpha}$ and apply (\ref {cf3}).

Notice that,   knowing  coefficients $C_{\alpha}$ for primary fields, we can express these coefficients for descendants. This allows us to rewrite (\ref{ope}) as a sum over primaries. 

We have defined the correlation functions on $\mathbb{R}^d.$  In a very similar way, one can define correlation functions on $S^d$ and find their relation to correlation functions on $\mathbb{R}^d$ using the fact that expressions $\psi_n(B_1, ...,B_n, h_1,...,h_n)$ are conformally invariant. Notice, however, that there exists no standard parameterizaion of a ball in $S^d$, therefore the correlation functions on $S^d$ depend not only on the points $x_1,...,x_n\in S^d$, but also on some additional data (for example, one can fix orthogonal frames at these points).
\section{AdS/CT dictionary.}

We identified   the group of conformal transformations of $S^d$ with the group of isometries of hyperbolic space $H^{d+1}.$  The Lie algebra $so(1, d+1)$ of this group acts on the space of boundary states. We identify the spaces of boundary states in CT and in AdS; they carry the same representation of $so(1,d+1).$

Let us discuss the interpretation of the subsemigroup $\mathcal {L}$ in AdS.  One can check directly that the generator of this semigroup, the dilation $S$,
in the language of the hyperboloid $x_0^2-...-x^2_{d+1}=1$ can be interpreted as "rotation" in the plane $(x_0, x_{d+1})$, i.e. as the vector field (infinitesimal transformation)
$$ \hat S= x_0\frac{\partial}{\partial x_{d+1}}+x_{d+1}\frac{\partial}{\partial x_0}.$$
This can be proven without calculations: we should look at geometric properties  of these transformations. In particular, it is clear that $\hat S$ transforms into itself the straight line in $H^{d+1}$ specified  by  the equations $x_1=...=x_d=0.$
This means that the corresponding transformation of the  ideal sphere should have two fixed points ; this is true for dilation $S.$

One can introduce coordinates $\tau,\rho, \Omega_i$ on hyperbolic space using the formulas
\begin{equation}
\label{h}
x_0=\frac{\cosh\tau}{\cos\rho},\\
\quad x_{d+1}=\frac{\sinh\tau}{\cos\rho},\\
\quad x_i=\tan\rho\Omega_i.
\end{equation}
In these coordinates $\hat S=\frac{\partial}{\partial \tau}.$
One can say that $\tau$ plays the role of (imaginary) time and the dilation in CT corresponds to the time translation in AdS. Hence scaling states correspond to stationary states in AdS, anomalous dimensions to energy levels.  Representations of  $so(1, d+1), $ generated by primary states  correspond to particle multiplets. In particular, the conserved energy-momentum tensor corresponds to  the graviton , because both of them are related to the same representation of $so(1, d+1).$  This justifies our statement that CT has conserved energy-momentum tensor (is a CFT) iff the AdS theory has the graviton in its spectrum. Conserved currents correspond to gauge particles.  (See \cite {KAP} for more detail).

Notice that our axioms of CT are not satisfactory in dimension 2. However, if we add to them the existence of conserved energy-momentum tensor we obtain two-dimensional CFT at genus zero (it is not clear whether we have modular invariance).  The energy-momentum tensor is not a primary  field in the standard definition of two-dimensional CFT, but it is a primary field in our definition; it can be considered as highest weight vector of some representation of $so(1,3).$ There are no propagating gravitons in three-dimensional gravity, however, we can define a graviton in three dimensions as a state that transforms according to the same representation of $so(1,3)$ as energy-momentum tensor in two dimensions. Then we can claim  that a weakly local field theory on $H^3$ containing graviton  induces genus zero two-dimensional CFT .
 
Let us  give geometric interpretation of the semigroup $\mathcal{L}$ in hyperbolic space. Recall that in $\mathbb{R}^d$ and in $S^d$ this semigroup is specified by the family of balls sitting inside a fixed ball and having common center. In hyperbolic space we have instead a family of half-spaces sitting inside a fixed half-space and orthogonal to a fixed straight line.
(Saying that the half-space is orthogonal to a straight line we have in mind that the bounding hyperplane is orthogonal to this line.) This statement will be used later in the proof of formula (\ref {ac}) . To prove the statement,  we recall that in coordinates $\tau, \rho, \Omega_i$  the transformations of the semigroup $\mathcal {L}$ are imaginary time translations $\tau\to \tau +const.$ This gives us an obvious  example of  the embedding of $\mathcal {L}$ in the   hyperbolic space $\mathcal {M}_2$   by half-spaces $\tau\leq const$ embedded in the half-space $\tau\leq 0$ (such a half-space together with half-space $\tau \geq 0$ determines a point of   $\mathcal {M}_2$.) It is clear that in this example half-spaces are orthogonal to the line $\rho=0, \Omega_i=0;$  we can say that $\cal L$ consists of shifts along this line. All other examples are obtained from this one by isometries (the group ${\rm Conf}_d$ acts on the space of straight lines transitively). Notice that to give the geometric interpretation of $\cal L$ we should fix not only half-spaces, but also their parameterizations; the coordinate description gives us the parameterizations we need.

Let us express the partition functions $\psi_n ( \Gamma_1,...\Gamma_n,h_1, ..., h_n)$ on the AdS side in terms of correlation functions of CT. By definition, these functions coincide with partition functions
$\psi_n ( B_1,..., B_n,h_1, ..., h_n)$ of CT theory (here $B_i$ are round balls corresponding to half-spaces $\Gamma_i$). Therefore it is clear that the expression in terms of correlation functions exists.  To describe this expression in more detail, we fix a point $O$ of hyperbolic space and draw a straight line starting at $O$ and going in the direction to $\Gamma _i$; we assume that this line is orthogonal  to the hyperplane bounding $\Gamma_i.$
We denote the ideal point of this line by $x_i.$ Then we can prove that
\begin{equation}
\label{ac}
\psi_n ( \Gamma_1,...\Gamma_n,h_1, ..., h_n)=e^{-\sum \rho_i\Delta_i}<\hat h_1(x_1) ...\hat h_n(x_n)>
\end{equation}
where $<\hat h_1(x_1) ...\hat h_n(x_n)>$ stands for correlation function on the sphere $S^d.$ We assume here that $h_i$ are scaling states with anomalous dimensions $\Delta_i$. The distance between $O$ and the hyperplane bounding $\Gamma_i$ is denoted by $\rho_i$; this distance can be positive or negative. 

To prove (\ref {ac}) we use the identification of $\mathcal{L}$ with family of half-spaces orthogonal to a fixed straight line. The formula follows immediately from (\ref{cf3}) and this identification.
 One can say it is a hyperbolic version of (\ref {cf3}).
 
 Notice that  the correlation function on the sphere $S^d$ entering the RHS of (\ref {ac}) depends not only on the points $x_1,...,x_n$, but also on some additional data (on orthogonal frames at these points); these data are specified by the parameterizations of the half-spaces $\Gamma_i.$

One can drop the assumption that $h_i$ are scaling states, then  (\ref {ac}) takes the form
\begin {equation}
\label{acc}
\psi_n ( \Gamma_1,...\Gamma_n,h_1, ..., h_n)=<\widehat  {e^{-\rho_1S}h_1}(x_1) ...\widehat {e^{-\rho_nS}h_n}(x_n)>
\end{equation}

Notice that we can take $\rho_i\to \infty$ in (\ref {ac}), then in the functional integral for $\psi_n$ we integrate fields defined on the whole hyperbolic space except "small" domains around $x_i$. (These domains are small in the Poincar\'e ball, but in hyperbolic space they are half-spaces.)  The elements $h_1,...,h_n$ specify the boundary  conditions on the boundaries of these domains. In this form (\ref {ac}) is close, but not identical, to the formulas in GKPW dictionary \cite {2}, \cite {3}, \cite {HAR},\cite {KAP}.

 To relate (\ref {ac}) to formulas in BDHM dictionary \cite {BD}, \cite {HAR}  one should calculate $<\phi (z), h;\Gamma>$ defined  as a partition function on half-space $\Gamma $ with boundary condition $h$ and  with insertion of the field $\phi$ at the point $z\in \Gamma$. Let us assume that the distance of $z$ from the point $O$ is equal to $r=r(z)$, the distance of    $\Gamma_{\rho}$ from the point $O$ is equal to $\rho$ and  $\Gamma_{\rho}$ is obtained from $\Gamma_0$ by means of a shift along the straight line connecting $O$ and $z$. (All $\Gamma_{\rho}$ are orthogonal to this line). We can consider $<\phi (z), h;\Gamma_{\rho}>$  as an ${\cal H}^*$ - valued function of $r$ and $\rho$ ( a linear functional on $\cal H$), but we will consider it as $\cal H$-valued function $F(r,\rho)$. (An inner product in $\cal H$ specifies an embedding of $\cal H$ into ${\cal H}^*$;  we identify $\cal H$ with the image of this embedding and  assume that $<\phi (z), h;\Gamma>$ lies in this image .) One can represent this function  in the form  
\begin{equation}
\label {g}
F(r,\rho)=e^{(\rho-r) S}h(\phi).
\end{equation}
 (Due to invariance with respect to isometries the function  $F(r, \rho)$ depends only on the difference $r-\rho$. From the other side it follows from the gluing formula that  $F(r,\rho')$ can be obtained from $F(r,\rho)$ by means of action of the operator $e^{(\rho'-\rho)S}$.)

Now we can calculate the correlation function $<\phi_1(z_1)...\phi_n(z_n)>$ obtained by insertion of the fields $\phi_1,...,\phi_n$ at the points $z_1,...,z_n$ of hyperbolic space in terms of correlation functions of CT on $S^d$. We assume that there exist non-overlapping half-spaces $\Gamma_i$ such that $z_i\in \Gamma_i$ and $\Gamma_i$ is orthogonal to the straight line connecting  $O$ and $z_i$. ( It is easy to get rid of this assumption.)Then the application of the gluing formula allows us to express the correlation function in terms of  $<\phi_i (z), h_i;\Gamma_i>$ and
$\psi_n ( \Gamma_1,...\Gamma_n,h_1, ..., h_n).$ Using (\ref{g}) and (\ref {acc}) we obtain
\begin{equation}
\label {gg}
<\phi_1(z_1)...\phi_n(z_n)>=<\widehat  {e^{-r_1S}h(\phi_1)}(x_1) ...\widehat {e^{-r_nS}h(\phi_n)}(x_n)>
\end{equation}
where $r_i$ is the distance from $O$ to $z_i.$ This formula generalizes the formulas of \cite {HK}.

The BDHM dictionary is based on the consideration of asymptotic behavior of the LHS in (\ref {gg}) as $r_i\to \infty$; we see that this behavior is governed by the homogeneous part of $h(\phi_i)$
having the minimal anomalous dimension; we denote this field by ${\bf{h}}(\phi_i)$ and the corresponding dimension by $\Delta_i$. Then (\ref{gg}) gives the asymptotic behavior of the LHS:
\begin{equation}
\label{ggg}
<\phi_1(z_1)...\phi_n(z_n)>\approx e^{-\sum r_i\Delta_i}<\widehat  {{\bf{h}}_1(\phi_1)}(x_1) ...\widehat{{\bf{ h}}_n(\phi_n)} (x_n)>.
\end{equation}
Formula (\ref {ac}) can be used in both directions: from CT to AdS or from AdS to CT. However, if we want to find the CT corresponding to a given theory on AdS it is better  to use different 
techniques.  Namely, one should take the domain bounded by two hyperplanes orthogonal to the fixed straight line and an isometric map of one hyperplane onto another hyperplane. We  construct  a non-compact  hyperbolic manifold  using the isometry to identify the hyperplanes. It is easy to express the partition function on this manifold (depending on the distance between the hyperplanes and on the element of $SO(1,d)$ specifying the isometry) in terms  of  the representation of $so(1,d+1)$ in the space of boundary states.  Conversely, knowing the partition function we can  get the information about this representation (that is the same in AdS and in CT). 
\section {Minkowski space}
We have worked in Euclidean setting ; it is not clear how to formulate similar axioms of CT in Minkowski space (or, better, in its compactification $S^{d-1}\times S^1$ or, even better, in the universal cover of the compactification ). This is an interesting problem. However, it is important to emphasize that the formula (\ref {gg}) can be analytically continued to Minkowski setting. (Notice, that the RHS of this formula is expressed in terms correlation functions on the sphere $S^d$; one should express these functions in terms of correlation functions on Euclidean space and then analytically continue to Minkowski space. The LHS  will give correlation functions on Lorentzian AdS.

The formula (\ref {gg}) allows us to apply the general theory of the paper \cite {SA} to quantum field theories on Lorentzian $(d+1)$-dimensional AdS . 

Notice that the starting point of  \cite {SA} is an algebra of observables $\cal A$  equipped with the action of commutative Lie group $T$.The elements of this Lie group are denoted by $(t,\vec x)$ and the corresponding automorphisms by $\alpha (t, \vec x)$; if $A\in\cal A$ we can consider a "field " $A(t,x)=\alpha (t,\vec x)A$. (Here $t\in \mathbb{R},\vec x\in \mathbb{R}^{d-1}$, the automorphisms $\alpha$ can be interpreted as time and space translations. The algebra $\cal A$ should be equipped with involution  $^*$, automorphisms commute with the involution.)  The state $\omega$ on the algebra $\cal A$ is defined as a linear functional $\omega$ obeying $\omega(1)=1, \omega (A^*A)\geq 0$. If the state $\omega$ is translationally invariant we can use the GNS (Gelfand-Naimark-Segal) construction to define  a (pre) Hilbert space $\cal E$  with action of the algebra $\cal A$ and the group $T$; the vector $\Omega$ corresponding to the state $\omega$ is annihilated by the generators of $T$ (by energy and momentum operators). If the energy operator is non-negative one says that  $\Omega$ is a physical vacuum.  We say that the theory is asymptotically commutative if the 
$||[\hat A(t,\vec x),\hat B]||$ tends to zero for large $\vec x$ and is polynomially bounded with respect to $t$ ( More precisely, we should require that $\int d\vec x ||[\hat A(t,\vec x),\hat B]||<g(t)$ where $g(t)$
is a polynomial. Operators $\hat A, \hat B$ in $\cal E$ correspond to observables $A,B\in \cal A$.) 

It was proven in \cite {SA} that  starting with  asymptotically commutative theory one can define scattering in $d$-dimensional space-time .  For a local field theory on $(d+1)$ -dimensional Lorentzian AdS one can construct the algebra of observables $\cal A$ using smeared fields.  The group ${\rm Conf}_d$ acting on this algebra contains a $d$-dimensional commutative subgroup $T$ generated by $P_{\mu}$ . (Notice that this subgroup cannot be extended to $(d+1)$-dimensional commutative subgroup, hence from the viewpoint   of \cite {SA}  $(d+1)$- dimensional AdS should be related to $d$-dimensional scattering.) It seems that using (\ref {gg}) one can prove asymptotic commutativity of the theory at hand; moreover, it seems that the same formula implies the coincidence of the scattering in the asymptotically commutative theory on $(d+1)$-dimensional theory on AdS and the scattering of the corresponding CT.

\section {Unitary theories}
It is well known that  unitarity in Minkowski space is equivalent to reflection positivity in the Euclidean approach \cite {OS}. It was proven in \cite {BH} that similarly unitarity in AdS is equivalent to reflection positivity in Euclidean AdS (in hyperbolic space). In this section we give a definition of reflection positivity in our setting.
The relation between reflection positivity in AdS and in CT  follows easily from this definition.
Let us fix a conformal $(d-1)$-dimensional sphere $S^{d-1}$ in $S^d$ or in $\mathbb{R}^d$. We say that
a conformal map $R$ is a reflection with respect to this sphere if it leaves all points of this sphere intact
(if the sphere is a hyperplane in $\mathbb{R}^d, $ this is an ordinary reflection, otherwise this is an inversion). 
The map $R$ induces a transformation $h\to h^*$ of the space of states $\mathcal{H}$
(we use this notation, because in the language of state-operator correspondence  the operator $\hat h^*(x)$ is adjoint to $\hat h(x)$).

The reflection positivity condition can be written in the form\begin{equation}
\label{rp}
\psi_2(R(B),B,h^*,h)\geq 0
\end{equation} 
where $B$ denotes a ball inside the fixed sphere. In more general form this condition can be written in the following way
\begin{equation}
\label{rp1}
\psi_{2n}(R(B_n),...,R(B_1),B_1,...,B_n,h^*_n,...,h^*_1,h_1,...,h_n)\geq 0
\end{equation}
where $B_1, ...,B_n$ are non-overlapping balls  inside the fixed sphere.

It is obvious that the reflection positivity condition in CT is equivalent to a similar condition in Euclidean AdS (in hyperbolic space). Instead of fixed sphere and reflection with respect to this sphere we should talk about fixed hyperplane and reflection with respect to this hyperplane, instead of balls we should consider half-spaces.

It seems that it is possible to check that the correlation functions in CT with reflection positivity property satisfy all axioms for Schwinger functions (Euclidean Green functions) of  unitary 
conformal  field theory in the sense of \cite {LM}.

\section {de Sitter space}
 
 One can construct the spaces $\mathcal{M}_n$   in the framework of
  de Sitter space, however, it is not clear that this construction can be used to derive the dS/CFT correspondence \cite {STR}.
  
   Both hyperbolic space and de Sitter space can be defined by the equation 
  $$<x,x>=c$$
  in the space $\mathbb{R}^{d+2}$. (Here $<x,x>=x_0^2-x_1^2-...-x_{d+1}^2$, for hyperbolic space $c=1$, for de Sitter space $c=-1.$) Half-spaces in the hyperbolic space can be specified by the formula $<a,x>\geq 0$ where $<a,a>< 0.$ We can use the same formula to define half-spaces in de Sitter space. Then we have one-to-one correspondence between half-spaces in these two spaces; this  leads to the construction of   $\mathcal{M}_n$ in terms of de Sitter space. (To construct the correspondence between parameterized half-spaces we should notice that the correspondence between half-spaces commutes with the action of ${\rm Conf}_d$.)  More precisely, in de Sitter space  we should  work with collections of half-spaces satisfying the condition that the corresponding  hyperplanes have compact intersections pairwise. The space of these collections is not precisely ${\cal M}_n$ , but  the closure of it is ${\cal M}_n$. This follows from the remark that    non-intersecting, but not parallel hyperplanes  in hyperbolic space correspond to hyperplanes on de Sitter space that have compact intersections (if the intersection is compact it is homeomorphic to a sphere).  Alas, one cannot directly apply the construction of Section 3 to the de Sitter space.

\vskip .1in
{\bf Acknowledgements} It is a pleasure to thank C. Beny, W. Chemissany .D.Fuchs, S. Gukov, V. Hubeny, M. Kapovich, M. Movshev, A. Polyakov, M.Rangamani,  J. Stasheff  and A. Tseytlin for help.


\bigskip
\footnotesize

A. Schwarz, \textsc{Department of Mathematics, University of California, Davis, CA 95616, USA} \par \nopagebreak \textit{E-mail address:} \texttt{schwarz@math.ucdavis.edu}

\end{document}